\documentclass[12pt,a4paper]{article}
\usepackage{xcolor}
\usepackage{jheppub}
\usepackage{epstopdf}
\usepackage{graphicx}
\usepackage{epsfig}
\usepackage{dcolumn}  
\usepackage{bm}    
\usepackage{amssymb} 
\usepackage{amsmath,bm}
\usepackage{amsfonts}    
\usepackage{slashed}  
\usepackage{youngtab}
\usepackage[mathscr]{euscript}
\usepackage{epsfig}
\usepackage{verbatim}
\newdimen\tableauside\tableauside=1.0ex
\newdimen\tableaurule\tableaurule=0.4pt
\newdimen\tableaustep
\def\phantomhrule#1{\hbox{\vbox to0pt{\hrule height\tableaurule width#1\vss}}}
\def\phantomvrule#1{\vbox{\hbox to0pt{\vrule width\tableaurule height#1\hss}}}
\def\sqr{\vbox{%
		\phantomhrule\tableaustep
		\hbox{\phantomvrule\tableaustep\kern\tableaustep\phantomvrule\tableaustep}%
		\hbox{\vbox{\phantomhrule\tableauside}\kern-\tableaurule}}}
\def\squares#1{\hbox{\count0=#1\noindent\loop\sqr
		\advance\count0 by-1 \ifnum\count0>0\repeat}}
\def\tableau#1{\vcenter{\offinterlineskip
		\tableaustep=\tableauside\advance\tableaustep by-\tableaurule
		\kern\normallineskip\hbox
		{\kern\normallineskip\vbox
			{\gettableau#1 0 }%
			\kern\normallineskip\kern\tableaurule}%
		\kern\normallineskip\kern\tableaurule}}
\def\gettableau#1 {\ifnum#1=0\let\next=\null\else
	\squares{#1}\let\next=\gettableau\fi\next}

\newcommand{\be}{ \begin{equation}}
\newcommand{\ee}{\end{equation}}
\newcommand{\bea}[1]{\begin{eqnarray}\label{#1} }
\newcommand{\eea}{\end{eqnarray}}

\def\ZZZ{{\hskip-3pt\hbox{ Z\kern-1.6mm Z}}}
\def\zzz{{\hskip-3pt\hbox{ z\kern-1mm z}}}

\def\one{{\hbox{ 1\kern-.8mm l}}}
\def\zero{{\hbox{ 0\kern-1.5mm 0}}}

\title{Stringy ${\rm AdS}_3$  from the Worldsheet}

\author{Matthias R.\ Gaberdiel$^a$,  Rajesh Gopakumar$^b$ and Chris Hull$^c$} 
\affiliation{$^a$ Institut f\"ur Theoretische Physik, ETH Zurich, \\
\hspace*{0.3cm}CH-8093 Z\"urich, Switzerland}
\affiliation{$^b$ International Centre for Theoretical Sciences-TIFR, \\
\hspace*{0.3cm}Survey No. 151, Shivakote, Hesaraghatta Hobli, \\
\hspace*{0.3cm}Bengaluru North, India 560 089}
\affiliation{$^c$ The Blackett Laboratory, Imperial College London \\
\hspace*{0.3cm} Prince Consort Road London SW7 2AZ, U.K.}

\emailAdd{gaberdiel@itp.phys.ethz.ch, \\ rajesh.gopakumar@icts.res.in, c.hull@imperial.ac.uk}

\abstract{We investigate the behaviour  of the bosonic string on  ${\rm AdS}_3$ with H-flux at stringy scales, looking 
in particular for a  `tensionless' limit in which there are massless higher spin gauge fields.
We do this by
revisiting the physical spectrum of the  $\mathfrak{sl}(2,\mathbb{R})_k$ WZW model
and considering the limit in which $k$ becomes small.
At $k=3$ we find that there is an infinite stringy tower of massless higher spin fields which are part of 
a continuum of light states. This can be viewed as a novel tensionless limit, which appears to be distinct 
from that inferred from the symmetric orbifold description of superstring ${\rm AdS}_3$ vacua.}

\preprint{Imperial-TP-2017-CH-01}

\begin{document}

\maketitle

\makeatletter
\g@addto@macro\bfseries{\boldmath}
\makeatother
\section{Introduction}

Bosonic string theory on three-dimensional anti-de Sitter space AdS$_3$ with H-flux or the superstring on AdS$_3$ with pure 
NS-NS background flux has a description in terms of a
WZW model based on $\mathfrak{sl}(2,\mathbb{R})$ 
\cite{Maldacena:2000hw,Maldacena:2000kv,Maldacena:2001km}. 
The theory has two dimensionful parameters, the string tension $T$ and the AdS cosmological constant $\Lambda$. The dimensionless 
ratio $T/\Lambda$ is  proportional to the WZW level $k$, which is a continuous real parameter for 
$\mathfrak{sl}(2,\mathbb{R})$.
The spectrum contains massive higher spin states with mass scale set by the string tension, so it is natural to ask 
whether an interacting massless higher spin theory could arise from a tensionless limit of the string theory in AdS
\cite{Sundborg:2000wp,Witten,Mikhailov:2002bp}. The naive tensionless string limit would be to take $k\to 0$, but it is 
also possible interesting behaviour could emerge as $k$ approaches a (small) critical value. 
In \cite{Lindstrom:2003mg}, the limit $k\to 2$ was considered, corresponding to the critical level 
of the $\mathfrak{sl}(2,\mathbb{R})$ affine algebra, and the possibility of massless higher spins arising in this limit was discussed. 

Our purpose in this note is to 
examine the space-time spectrum of strings on AdS$_3$ (tensored with an internal CFT so as to give a critical string theory) as a function of  
$k$. In particular, we analyse
whether the spectrum contains massless higher spin states at some critical value of $k$, corresponding to a 
tensionless limit of string theory on this background. We find that, while the $k\to 2 $ limit appears problematic, $k=3$ gives a spectrum with 
an infinite number of massless higher spin fields. 

These massless higher spin fields are the lowest members of a continuum of physical states. This continuum arises from the continuous 
representations of $\mathfrak{sl}(2,\mathbb{R})$ in the so-called spectrally flowed sector (with `winding number' $w=1$). Physically, for large $k$
these 
correspond to modes of a \lq long string' 
near the boundary 
of AdS  \cite{Maldacena:1998uz,Seiberg:1999xz, Maldacena:2000hw}. The tension of these strings tends to zero in the limit $k\to 3$, so the limit 
we are considering can be thought of as that in which long fundamental strings become effectively
tensionless due to a cancellation between the contributions of the H-flux and their actual tension. Since these are strings 
stretched near the boundary of AdS, they have radial momentum excitations which give rise to the continuum. 
Roughly speaking, as $k$ becomes small, the curvature of AdS approaches the string scale and massive states become 
effectively localised toward the centre of AdS and only massless states can approach the boundary, so the only way 
states associated with the long string can stay  near the boundary is for the string to become  tensionless.

This is somewhat different from the tensionless limit that one sees in other AdS vacua, in which the actual tension of the 
fundamental string becomes small (in comparison to the AdS scale). In AdS$_{d}$ with $d>3$ 
the tensionless limit corresponds to a free gauge theory, which has additional conserved currents, thus implying an 
enhanced gauge symmetry in the bulk. It corresponds to the leading Regge trajectory of stringy states becoming massless. 
This is the subsector which is believed to be described by a Vasiliev theory \cite{Vasiliev:2003ev} of higher spin gauge fields 
\cite{Sundborg:2000wp,Witten,Mikhailov:2002bp}. 

For the AdS$_3$ superstring (with R-R three-form flux), the tensionless limit is that in which the dual CFT$_2$ is the free symmetric orbifold theory. There, in contrast 
to the higher dimensional cases, we have a stringy tower of massless higher spin fields rather than a single Regge trajectory. 
This tensionless limit has been studied recently in \cite{Gaberdiel:2014cha,Gaberdiel:2015mra,Gaberdiel:2015wpo}, 
and a large underlying  unbroken symmetry, dubbed the ``higher spin square", has been uncovered. The tensionless limit 
discussed here appears to be a more singular one in that the continuum of states would lead to a similar continuum in the dual 
CFT signalling, perhaps, a noncompact direction.  However, it shares  with the regular tensionless limit of the 
AdS$_3$ superstring the feature of having a stringy tower of massless states. We comment more on these similarities and differences 
in the discussion section. Before that, in the next section, we briefly review the bosonic string spectrum of 
Maldacena-Ooguri  \cite{Maldacena:2000hw,Maldacena:2000kv,Maldacena:2001km}, and then go on to examine the special features at small $k$, including the new tensionless limit.

\section{The worldsheet description of the spectrum}

The ${\rm SL}(2,\mathbb{R})$ group manifold is a three-dimensional hyperboloid with  periodic time,  giving rise to 
a discrete energy spectrum. We will work with AdS$_3$, which is the universal covering space of this hyperboloid, and 
which has non-compact time so that the energy spectrum  is not forced to be discrete.
We will investigate the critical bosonic string theory formulated in terms of the WZW model of level $k$ on 
AdS$_3$, combined with an `internal' CFT to give a CFT with total central charge $c=26$.
This theory has been studied extensively in \cite{Maldacena:2000hw,Maldacena:2000kv,Maldacena:2001km,%
Hwang:1990aq,Evans:1998qu,Giveon:1998ns,Israel:2003ry,Raju:2007uj},
and we will make extensive use of  the results of \cite{Maldacena:2000hw}.

\subsection{Conventions and setup}

Let us begin by fixing our notation and conventions, mostly following \cite{Maldacena:2000hw}. 
For general level $k$, the commutation relations of the $\mathfrak{sl}(2,\mathbb{R})$ algebra are 
\begin{eqnarray}
{}[J^3_m,J^\pm_n] & = &  \pm\, J^\pm_{m+n} \\
{}[J^+_m,J^-_n] & = & - 2\, J^3_{m+n} + k \, m \, \delta_{m,-n} \\
{}[J^3_m,J^3_n] & = & - \tfrac{k}{2} \, m\, \delta_{m,-n} \ . 
\end{eqnarray}
For $k\neq 2$, the Sugawara construction is well defined, 
\begin{equation}\label{vir1}
\begin{array}{rcl}
L_n & = &{\displaystyle \frac{1}{k-2} \, \sum_{m=0}^{\infty} \bigl( J^+_{n-m}  J^-_m + J^-_{n-m} J^+_m - 2 J^3_{n-m} J^3_m \bigr) \qquad
(n\neq 0) }  \\[4pt]
L_0 & = & {\displaystyle \frac{1}{k-2} \, \Bigl[ \frac{1}{2} \bigl( J^+_0 J^-_0 + J^-_0 J^+_0 \bigr) - J^3_0 J^3_0 + 
\sum_{m=1}^{\infty} \bigl( J^+_{-m}  J^-_m + J^-_{-m} J^+_m - 2 J^3_{-m} J^3_m \bigr) \Bigr]  \ ,}
\end{array}
\end{equation}
and the resulting Virasoro generators satisfy the commutation relations
\begin{eqnarray}
{}[L_m,L_n] & = & (m-n) \, L_{m+n} + \frac{c}{12}\, m\, (m^2-1)\, \delta_{m,-n} \\
{}[L_m,J^a_n ] & = & - n  \,J^a_{m+n} \ , 
\end{eqnarray}
where the central charge takes the value
\be
c = \frac{3\, k}{k-2} \ . 
\ee
For this to satisfy $c\le 26$ requires $k\ge 52/23$, which in particular excludes $k=2$, so that the $k=2$ WZW model 
cannot be part of a conventional critical string theory.

The critical level is $k_{\rm crit}=2$ since then the  Virasoro generators (\ref{vir1}) 
are ill-defined.   For this reason,  it will be useful to define also the generators $l_n\equiv (k-2)\, L_n$, i.e.\ 
\begin{eqnarray}
l_n & = &  \sum_{m=0}^{\infty} \bigl( J^+_{n-m}  J^-_m + J^-_{n-m} J^+_m - 2 J^3_{n-m} J^3_m \bigr) \qquad
(n\neq 0)  \label{virp1} \\
l_0 & = &  \Bigl[ \frac{1}{2} \bigl( J^+_0 J^-_0 + J^-_0 J^+_0 \bigr) - J^3_0 J^3_0 + 
\sum_{m=1}^{\infty} \bigl( J^+_{-m}  J^-_m + J^-_{-m} J^+_m - 2 J^3_{-m} J^3_m \bigr) \Bigr] \ .  \label{virp2} 
\end{eqnarray}
They satisfy the commutation relations 
\begin{eqnarray}
{}[l_m,l_n] & = & (k-2)\, (m-n) \, l_{m+n} + \bigl(3 \, k\,  (k-2) \bigr) \, m\, (m^2-1)\, \delta_{m,-n} \\
{}[l_m,J^a_n ] & = & - (k-2)\, n  \,J^a_{m+n} \ .
\end{eqnarray}
In particular, for $k=2$, the $l_m$ generators are central, i.e.\ they commute both with the current modes $J^a_n$,
and with themselves. The existence of these commuting charges suggests that the theory at $k=2$
could be integrable. 

At $k=2$, the  conventional Virasoro constraints involving $L_n$ cannot be imposed, giving the possibility of evading 
the exclusion of $k=2$ \cite{Lindstrom:2003mg}. Moreover, at the critical level of $k=2$, affine representations have 
a large number of null states of the form \cite{Feigin}  
\be
l_{n_1} l_{n_2} \dots l_{n_p}  | \phi \rangle \ ,
\ee
where  $n_i>0$ are positive integers and $| \phi \rangle$ is any state satisfying
$
J^a_n | \phi \rangle=0
$
for $n>0$.
One may take this as evidence that also the space-time theory will develop a large
gauge symmetry at $k=2$, and a natural idea could be that this gauge symmetry is related to a
higher spin theory or a tensionless string theory \cite{Lindstrom:2003mg}.\footnote{Similar ideas
have been pursued  in \cite{Sagnotti:2003qa,Bakas:2004jq}, see also \cite{Bagchi:2015nca}.}
However,   this expectation
does not seem to be born out by a careful analysis of the spectrum.
Instead, it seems that something special happens for $k=3$. 

\subsection{Affine representations}\label{sec:affine}

We shall be interested in representations of the affine $\mathfrak{sl}(2,\mathbb{R})$ algebra based on
Virasoro highest weight representations (which, however,  need not be affine highest weight). 
In Section~\ref{sec:specflow} we will  look also at more general classes of representations that arise
as the spectrally flowed images of these representations and which are not even Virasoro highest weight 
\cite{Maldacena:2000hw}; however, let us first consider this simpler setting.

We will denote the Virasoro highest weight states by $|\hat{c},m\rangle$, where $\hat{c}$ is the eigenvalue of 
the quadratic Casimir $C$ of $\mathfrak{sl}(2,\mathbb{R})$ 
\be
C = \frac{1}{2} \bigl( J^+_0 J^-_0 + J^-_0 J^+_0 \bigr) - J^3_0 J^3_0 \ ,
\ee
while $m$ is the eigenvalue of $J^3_0$, i.e.\
\be
C |\hat{c},m\rangle  = \hat{c}\,  |\hat{c},m\rangle \ , \qquad 
J^3_0 \, |\hat{c},m\rangle  = m\,  |\hat{c},m\rangle \ . 
\ee
Here `Virasoro highest weight' means that 
\be
J^a_n |\hat{c},m\rangle = 0 \ , \qquad \hbox{for $n>0$}
\ee
since this implies, in particular, that $L_n |\hat{c},m\rangle = 0$ for $n>0$; however, we do not assume
that any of the states is annihilated by $J^+_0$ (or indeed $J^-_0$), i.e.\ we do not assume  that it is `affine highest 
(or lowest) weight'. 

There are five classes of unitary affine representations, and of these only two play a role in the spectrum of 
\cite{Maldacena:2000hw}. 
The two classes of (Virasoro) highest weight representations that appear in the spectrum of 
\cite{Maldacena:2000hw} are: (i) the discrete lowest weight representations ${\cal D}^+_j$ for which 
$m= j, j+1, j+2, \ldots$ and $\hat{c} = \hat{c}(j) \equiv -j (j-1)$ (with $j$ positive), and (ii) the continuous 
representations ${\cal C}(p,\alpha)$ labelled by $p,\alpha\in\mathbb{R}$ for which the 
$J^3_0$ spectrum is unbounded both from below and above --- $m$ takes the values
$m\in\alpha+\mathbb{Z}$ --- and $\hat{c} =  \frac{1}{4} + p^2$. Without loss of generality, the 
real parameter $\alpha$ can be taken to be in the range $0\le \alpha <1$.
For $k\neq 2$, we can immediately read off the conformal dimension of the ground states;
indeed it follows from (\ref{vir1}) that 
\be
L_0 \, |\hat{c},m\rangle = \frac{\hat{c}}{k-2} \, |\hat{c},m\rangle \quad \Longrightarrow \ h_{\rm{SL(2,\mathbb{R})}} = \frac{\hat{c}}{k-2} \ . 
\ee
The full affine representation is then obtained from these Virasoro highest weight states by 
the action of the negative current modes, modulo null vectors. The physical states are 
characterised by the condition that they are annihilated by $L_n$ with $n>0$. In addition, they have
to satisfy the mass-shell condition 
\be\label{massshell} 
 h_{\rm{SL(2,\mathbb{R})}} + h_{\rm{int}} + N= \frac{\hat{c}}{k-2} + h_{\rm{int}} + N = 1 \ , 
\ee
where $\hat{c}$ is the value of the Casimir operator  $C$ on the ground states, while 
$h_{\rm{int}}$ is the conformal dimension of the internal CFT (making up the additional directions of the 
full string theory, in addition to ${\rm AdS}_3$), and $N$ is the excitation number of the state. In the following
we shall concentrate on the case $h_{\rm{int}}=0$, and take $N$ to arise entirely from $\mathfrak{sl}(2,\mathbb{R})$ excitations, 
i.e.\ we take the `internal' CFT to be in its trivial ground state. (It will become obvious from the following discussion that our 
conclusions will not be changed if we included excitations in the internal CFT.)

We can now relate these worldsheet quantities to spacetime variables. Physical states in AdS$_3$
fit into unitary representations of the anti-de Sitter group ${\rm SL}(2,\mathbb{R})_L \times {\rm SL}(2,\mathbb{R})_R$, 
and for lowest-weight representations the energy is bounded below.
The spacetime energy $E$ and spin $s$ of a state with $J_0^3$ eigenvalue $m$ and $\bar J_0^3$ eigenvalue $\bar m$
are
\be
E = m+\bar m \ , \qquad s = m- \bar m \ .
\ee
The AdS$_3$ 
mass is given by the quadratic Casimir and is defined by 
\be\label{dict}
m^2_{{\rm AdS}_3} = (E - |s|) ( E + |s| -2) \ . 
\ee
There is a Breitenlohner-Freedman (BF) bound $m^2_{{\rm AdS}_3}  \geq -(|s|-1)^2$.

Massless higher spin fields correspond to states for which $E=|s|$.
For a 
 lowest weight representation of ${\rm SL}(2,\mathbb{R})_L \times {\rm SL}(2,\mathbb{R})_R$ 
with lowest weights given by $j_L=m_L = (E+s)/2$ and $j_R=m_R = (E-s)/2$,
the representation becomes reducible if $E=|s|$; if this is the case the representation
contains a sub-representation of null states that describe longitudinal (pure gauge) modes which decouple. 
The corresponding spacetime fields are then massless higher spin gauge fields. 

The conformal dimension of the 
 spacetime 
 CFT, $h_{\rm CFT}$, is to be identified with the $J^3_0$ eigenvalue $m$, 
and likewise for the right-movers, i.e.\ $\bar{h}_{\rm CFT}=\bar{m}$, the eigenvalue of $\bar{J}^3_0$. 
The  Breitenlohner-Freedman (BF) bound 
follows from the requirement of 
unitarity of the dual CFT, i.e.\ $ h_{\rm CFT},\bar{h}_{\rm CFT} \geq 0$.

\subsection{Massless Higher Spin States: The spectrally unflowed case}

Returning to the worldsheet, let us now see whether there exist massless higher spin states in the spectrum for some (small) value of $k$. 
We will first discuss the so-called spectrally unflowed sector corresponding to the two classes of Virasoro highest weight representations 
of the previous subsection. We will discuss the spectrally flowed versions of these representations in the following subsection.

For the continuous representations,
the first term in (\ref{massshell}) is positive for $k>2$, and the only possible solution is $N=0$ (since $h_{\rm int} \geq 0$). 
These representations have spacetime energies unbounded from below --- they violate the BF bound. The corresponding 
spacetime state is a tachyon; such a state was to be expected for bosonic string theory. 

For the discrete (lowest weight) representations labelled by $j$, the Casimir $\hat{c} = \hat{c}(j) \equiv - j (j-1)$ is negative,
and there are more interesting solutions. First we solve (\ref{massshell}) for $j$ (as a function of $N$), to conclude that 
\be\label{jsol}
j =  \frac{1}{2} + \sqrt{\frac{1}{4} + (k-2) (h_{\rm{int}}+ N-1)} \ . 
\ee
The no-ghost theorem \cite{Hwang:1990aq,Evans:1998qu} implies that $j$ has to be bounded by 
\be\label{unitarity}
0 \leq j \leq \frac{k}{2} \ . 
\ee
Actually, in the analysis of \cite{Maldacena:2000hw} a somewhat stronger bound is imposed, namely
\be\label{MObound}
\frac{1}{2} < j < \frac{k-1}{2} \ . 
\ee
Here the reason for the lower bound $\frac{1}{2}<j$ comes from the requirement that the corresponding
ground state $\mathfrak{sl}(2,\mathbb{R})$ representation should lead to square-integrable wave-functions,\footnote{This conclusion 
may be relaxed by considering an alternate quantisation of modes with $0 < j < \frac{1}{2}$, see page 18 of  \cite{Maldacena:2000hw}.} 
and this, via spectral flow, then also leads to the stronger upper bound  $j < \frac{k-1}{2}$.
Furthermore, it seems from the analysis of  \cite{Maldacena:2000kv} that the 
MO-bound  (\ref{MObound}) is required for modular invariance. 

As we have seen,
 massless higher spin fields correspond to states for which $E=|s|$. In terms of the 
conformal dimensions of the spacetime CFT, this means that either $h_{\rm CFT}=0$ or $\bar{h}_{\rm CFT}=0$, so that the
operators describe chiral fields. 
Thus we need to investigate whether the physical string spectrum contains states with $h_{\rm CFT}=m=0$, say.

Given the ground state spin $j$ together with the excitation number $N$, the possible $\mathfrak{sl}(2,\mathbb{R})$ 
lowest weights $\hat{\jmath}$ lie in the range $j-N \leq \hat{\jmath} \leq j+N$ --- this just follows from the fact that the oscillators
sit in the adjoint representation of $\mathfrak{sl}(2,\mathbb{R})$, and tensoring with the adjoint representation
can change the spin at most by $\pm 1$. Thus the minimal value of $m$ that can appear is $m_{\rm min} = j -N$. 
The question is then whether $j-N=0$ is possible, provided that $j$ is given by (\ref{jsol}) and $j$ obeys the
unitarity bound (\ref{unitarity}) or the Maldacena-Ooguri bound (\ref{MObound}). [Since $h_{\rm CFT}=\hat{\jmath}$ is the conformal
dimension of the dual CFT, the no-ghost theorem (i.e.\ unitarity) guarantees that $h_{\rm CFT}\geq 0$ --- so the only way $h_{\rm CFT}=0$ can 
arise is for $h=j-N=0$.]

One solution is obvious: if we take $N=1$ (with $h_{\rm int}=0$), then $j=1$ and $j-N=0$. Tensoring this state together with the corresponding
right-mover with $\bar{\jmath}= j+N = 2$ gives then the helicity two component of the graviton.
This state is allowed by unitarity for $k\geq 2$, whereas the MO-bound 
(\ref{MObound}) only allows   the massless graviton for $k\geq 3$. 

In the following we want to study whether there are other solutions for $N\geq 2$, corresponding to higher spin. Going back
to the original equation (\ref{massshell}) for $h_{\rm int}=0$ and with $\hat{c}(j) = - j (j-1)$, we have to solve\footnote{It is clear that 
there can be no solutions with non-zero $h_{\rm int}$: if $h_{\rm int}>0$, the relevant $j$ in (\ref{jsol}) is larger than the value for 
$h_{\rm int}=0$, and hence $j-N$ cannot be zero --- unitarity of the dual CFT (which follows from the no-ghost theorem) implies 
that already  the original $j-N$ was non-negative, so increasing $j$ cannot make it zero.}
\be\label{eq0}
- \frac{j (j-1)}{k-2} + N = 1 \ ,
\ee
which we can rewrite as 
\be\label{eq1}
j^2 - j - (k-2) (N-1) = 0 \ . 
\ee
In order for $h=j-N=0$, this has to be solved for $j=N$. Provided that $N\geq 2$, plugging $j=N$ into
(\ref{eq1}) leads to 
\be\label{eq2}
N^2 - N - (k-2) (N-1) = 0 \qquad \Longrightarrow \qquad N= k - 2 \ . 
\ee
For $N=2,3,\ldots$, corresponding to $k=4,5,\ldots$, it then follows that 
\be
j = N = k-2 \geq \frac{k-1}{2} \ , 
\ee
in violation of the upper bound on $j$ from (\ref{MObound}).
We note in passing that if we were to relax the upper bound on $j$ to the unitary bound (\ref{unitarity}), then we have 
the solution $k=4$ with $N=2=j$. However, the corresponding state 
\be
J^{-}_{-1} J^-_{-1} |j=2 \rangle
\ee
is null since it is a descendant of the null-state $J^-_{-1} |j=2 \rangle$ at level $k=4$.

\subsection{The spectrally flowed representations}\label{sec:specflow}

It was seen in \cite{Maldacena:2000kv} that, as well as the positive energy representations discussed above, 
the formulation of string theory in AdS$_3$ also requires spectrally flowed representations that are not Virasoro highest weight. 
One way to describe the spectrally flowed representations is as follows. We start with
the vector space corresponding to a Virasoro highest weight representation of 
$\mathfrak{sl}(2,\mathbb{R})_k$ --- in our case, this will either be a lowest weight representation
${\cal D}^+_j$ or a continuous representation ${\cal C}(p,\alpha)$. On this vector
space we then define the action of $\mathfrak{sl}(2,\mathbb{R})_k$ and the Virasoro algebra by 
the action of the hatted operators defined via
\begin{eqnarray}
\hat{J}^\pm_n & = & J^\pm_{n\mp w} \nonumber \\[2pt]
\hat{J}^3_n & = & J^3_n + \tfrac{k}{2} w \delta_{n,0} \nonumber \\[2pt]
\hat{L}_n & = & L_n - w J^3_n - \tfrac{k}{4} w^2 \delta_{n,0} \ . \label{virtrans}
\end{eqnarray}
With respect to the hatted operators the vector space is then not a conventional (Virasoro) highest weight
representation. However, provided that $w>0$, the representation (with respect to the hatted modes)
consists of lowest weight representations of the global $\mathfrak{sl}(2,\mathbb{R})$ algebra --- this
is the case of physical interest, since then the energy spectrum in AdS$_3$ is bounded from below
(and so the spectrum of dimensions in
 the dual CFT will be bounded from below). 

With respect to this hatted action we  impose the usual mass-shell condition, which now takes the form
(from now on we are setting $h_{\rm int}=0$) 
\be\label{massflow}
\frac{\hat{c}}{k-2}  - w m -\frac{k}{4} w^2 +N = 1 \ , 
\ee
where $m$ is the $J^3_0$ eigenvalue of the state in question. Furthermore, the conformal dimension of the state
in the dual CFT is now $h_{\rm CFT}= m + \frac{k}{2} w$. As before the presence of a massless higher spin state requires 
$h_{\rm CFT}= m + \frac{k}{2} w=0$, i.e.\ that we take $m=-\frac{k}{2}w$. Plugging this into (\ref{massflow}) 
leads to 
\be\label{massflow1}
\frac{\hat{c}}{k-2}  + \frac{k}{4} w^2 + N  = 1 \ .
\ee
Let us now consider the two classes of representations in turn.  
For the discrete representations $\hat{c}=-j(j-1)$ and, since $j$ still has to satisfy the unitarity bound 
(\ref{unitarity}), we conclude that 
\be\label{ineq}
- \frac{j (j-1)}{k-2} \geq - \frac{k}{4} \ ,
\ee
and hence that 
\be
1 - N - \frac{k}{4} (w^2-1) \geq 0 \ . 
\ee
Thus the only cases to consider are $w=1$ with $N=0,1$. Furthermore, since $m=-\frac{k}{2}w = -\frac{k}{2}$, 
but $j\geq 0$ and the minimal value of $m$ is $j-N = -1$ (for $j=0,N=1$), the only possible solution is $k=2$, which is not allowed --- at $k=2$ the denominator in the stress energy tensor blows up.\footnote{In addition, $j=0$
is not allowed by the Maldacena-Ooguri bound, eq.~(\ref{MObound}).} Thus there are no massless
higher spin fields from the spectrally flowed discrete representations.
\medskip

For the continuous representations $\hat{c} = p^2 + \frac{1}{4}$. 
Since all three terms are non-negative, the only possible solution arises for $N=0$ and $w=1$, for which we obtain 
the equation
\be
p^2 + \frac{1}{4} = - \frac{k^2}{4} + \frac{3}{2}k  -2 \ .
\ee
It is not difficult to show that the only solution arises for $k=3$ and $p=0$. For these values, i.e.\
$k=3$, $p=0$, and $w=1$, the solution to the mass-shell condition (\ref{massflow}) is then 
\be
- m - \frac{1}{2} + N = 1 \ , 
\ee
which leads to $m = N - \frac{3}{2}$. Thus, in addition to the massless states, we get a whole tower of states (each at 
the bottom of a continuum labelled by $p$) with $h_{\rm CFT}=N$ and $\bar{h}_{\rm CFT} = \bar{N}$, where
we recall that $h_{\rm CFT}=m+ \frac{k}{2} w = m + \frac{3}{2}$, 
and similarly for the right-movers. To summarise, the infinite tower of massless higher spin fields are given by 
considering the continuous representation with $p=0$ in the $w=1$ sector at $k=3$, where we take 
\be
m = -\frac{3}{2} + N  \ , \qquad \bar{m} = -\frac{3}{2} + \bar{N} \ , 
\ee
with $N,\bar{N} \in \mathbb{N}_0$ and either $N=0$ or $\bar{N}=0$. 
Note that these states (for any $N$, $\bar{N}$), satisfy the level matching condition 
\cite{Maldacena:2000hw}, for non-zero spectral flow, which requires that $w(m-\bar{m})$ be an integer. 

We should also note that the graviton which was part of the $w=0$ discrete spectrum merges at this point with the 
continuum since it has $j=1 =\frac{(k-1)}{2}$ and thus can be identified with the state in the $w=1$ continuous 
representation with $N=0$, $\bar{N}=2$ (as well as the state where $N$ and  $\bar{N}$ are interchanged), cf.\
the  comments around page 25 of \cite{Maldacena:2000hw}. 
\section{Discussion}

We have found, from a close examination of the bosonic string spectrum on AdS$_3$ with H-flux, that there is a novel kind of 
tensionless limit at small $k$. This happens at $k=3$, rather than the critical value $k=2$ which one might have expected. 
In fact, as we have seen, according to the Maldacena-Ooguri bound (\ref{MObound}), the massless graviton (with $j=1$) is 
only part of the physical spectrum for $k\geq 3$. Therefore, the meaning of the theories with $2 \le k <3$ is somewhat unclear.  (For $k=2$, there is a massless spin-two state consistent with the unitarity bound (\ref{unitarity}), but it is not normalisable. It is possible that an alternative approach to the quantisation of the $k=2$ case, perhaps with a non-standard norm on the space of states, could be useful here.)

It will be interesting to study the physics of this tensionless limit further, and understand the similarities and 
differences with the limit corresponding to the symmetric orbifold point.  Seiberg and Witten
\cite{Seiberg:1999xz}, in work   preceding that of Maldacena and Ooguri, 
studied long fundamental strings in AdS$_3$ and argued that  a long string is effectively described, for $k>3$, by a Liouville
theory with background charge $Q$ and central charge $c=1+3Q^2$, where
\be 
Q= (k-3) \sqrt{ \frac 2 {k-2}} \ . 
\ee 
The Liouville field is associated with the distance of the long string from the boundary of AdS$_3$.
The Liouville theory  has a spectrum that is discrete up to a threshold energy 
\be
\Delta_0=\frac{Q^2}{8} =\frac{(k-3)^2}{4(k-2)} \ , 
\ee
and continuous above this, see eq.~(4.15) of \cite{Seiberg:1999xz}. 
This is exactly the behaviour that we find from the   microscopic analysis of Maldacena and Ooguri --- if 
we put $p=0$, $w=1$, and $N=0$,  and solve for $m$ from the on-shell condition (\ref{massflow}), we find that 
$h_{\rm{CFT}}=m+\frac{k}{2}= \frac{(k-3)^2}{4(k-2)}$.  
In the limit $k\to 3$, the threshold tends to zero and we obtain a continuous spectrum.
The discussion of Seiberg and Witten
\cite{Seiberg:1999xz} suggests that this continuous spectrum could be associated with a  limit of Liouville theory
in which the background charge $Q\to 0$ giving a $c=1$ CFT.
A natural hypothesis is that this could be a free scalar field, corresponding to an extra 
non-compact dimension emerging in this limit.\footnote{Note, however, that taking this limit directly in 
Liouville theory is tricky since  the theory is only conformally invariant if  the charge $Q$ is related to the 
parameter $\gamma$ in the exponential Liouville potential $\mu^2  e^{\gamma \phi}$ by
$Q= \gamma+ 2/\gamma$ (see \cite{Seiberg:1990eb} and references therein). Therefore, 
taking $Q < 2\sqrt{2}$ is not possible for real $\gamma$. An accurate comparison with a 
candidate dual non-compact CFT will therefore need to go beyond the effective Liouville 
description (which breaks down for $k\approx 8$) and be based on the microscopic analysis of \cite{Maldacena:2000kv}.}

Seiberg and Witten
\cite{Seiberg:1999xz} also considered the supersymmetric case, resulting from the near horizon limit of  a 
system of $Q_1$ fundamental strings inside $Q_5$ NS 5-branes, finding a similar behaviour with a 
threshold energy of 
\be
\Delta_0=\frac {Q^2}{8}  = \frac{(Q_5-1)^2}{4Q_5} \ , 
\ee
so that $Q_5\to 1$ is the limit in which $\Delta_0 \to 0$ and there are extra massless states.
Further, it is conjectured in \cite{Seiberg:1999xz} that the dual CFT  has a 
$(\mathbb{R}^4)^{Q_1}/S_{Q_1}$ factor. 
The spectrum of the superstring case is treated in more detail in \cite{FGJ}.

The appearance of a continuum in the dual CFT is also somewhat reminiscent 
of the light states in the vicinity of the (Vasiliev) tower of higher spin currents in the ${\cal W}_N$ minimal models in the 
large $N$ limit, see \cite{Gaberdiel:2010pz,Gaberdiel:2013cca}. It would be useful to see whether this can be made more precise. 

We have seen that there is a stringy tower of massless higher spin states that become massless in the limit $k\to 3$. One may 
ask about the unbroken gauge symmetry that they generate. Naively, it seems to be similar in structure to that of the Higher Spin Square 
\cite{Gaberdiel:2015mra,Gaberdiel:2015wpo} that is seen at the symmetric orbifold point. This would indeed be so if we can 
identify these states in the dual CFT with the single particle states in a symmetric product of free non-compact bosons 
of the kind mentioned above.

\section*{Acknowledgements}

We thank Kevin Ferreira, Juan Jottar, David Kutasov, Wei Li, Ulf Lindstrom, and Juan Maldacena 
for useful discussions. We thank the Galileo Galilei Institute for Theoretical Physics (GGI) for the 
hospitality, and INFN for partial support during the completion of this work, during the program 
``New Developments in AdS3/CFT2 Holography".  The work of MRG is supported in part 
by the NCCR SwissMAP, funded by the Swiss National Science Foundation. 
The  work of CH was supported in part by a grant from the Simons Foundation,  the EPSRC programme grant 
``New Geometric Structures from String Theory" EP/K034456/1 and the STFC  grant ST/L00044X/1.

\end{document}